# A COMBINED ON-LINE ACOUSTIC FLOWMETER AND FLUOROCARBON COOLANT MIXTURE ANALYZER FOR THE ATLAS SILICON TRACKER

A. Bitadze*
SUPA, School of Physics and Astronomy
University of Glasgow, Scotland
Email: a.bitadze@physics.gla.ac.uk

*On behalf of collaboration*:
R. Bates, Dept. Physics & Astronomy, University of Glasgow, G12 8QQ, Scotland, UK
M. Battistin, S. Berry, P. Bonneau, J. Botelho-Direito, B. DiGirolamo, J. Godlewski,
E. Perez-Rodriguez, L. Zwalinski, CERN, 1211 Geneva 23, Switzerland
N. Bousson, G. Hallewell, M. Mathieu, A. Rozanov, CPPM, F-13288 Marseille, France
G. Boyd, Dept. Physics & Astronomy, University of Oklahoma, Norman, OK 73019, USA
M. Doubek, V. Vacek, M. Vitek, Czech Technical University, 16607 Prague 6, Czech Republic
K. Egorov, Physics Department, Indiana University, Bloomington, IN 47405, USA
S. Katunin, B.P. Konstantinov Petersburg Nuclear Physics Institute, 188300 St. Petersburg, Russia
S. McMahon, STFC Rutherford Appleton Laboratory, Chilton, Didcot OX11 OQX, UK
K. Nagai, Graduate School of Pure and Applied Sciences, University of Tsukuba,
1-1-1 Tennodai, Tsukuba, Ibaraki 305-8577, Japan

## Abstract

An upgrade to the ATLAS silicon tracker cooling control system may require a change from $C_3F_8$ (octafluoro-propane) to a blend containing 10-30% of $C_2F_6$ (hexafluoro-ethane) to reduce the evaporation temperature and better protect the silicon from cumulative radiation damage with increasing LHC luminosity.

Central to this upgrade is a new acoustic instrument for the real-time measurement of the $C_3F_8/C_2F_6$ mixture ratio and flow. The instrument and its Supervisory, Control and Data Acquisition (SCADA) software are described in this paper.

The instrument has demonstrated a resolution of $3.10^{-3}$ for $C_3F_8/C_2F_6$ mixtures with ~20%$C_2F_6$, and flow resolution of 2% of full scale for mass flows up to $30 gs^{-1}$. In mixtures of widely-differing molecular weight (mw), higher mixture precision is possible: a sensitivity of $< 5.10^{-4}$ to leaks of $C_3F_8$ into the ATLAS pixel detector nitrogen envelope (mw difference 160) has been seen.

The instrument has many potential applications, including the analysis of mixtures of hydrocarbons, vapours for semi-conductor manufacture and anaesthesia.

# INTRODUCTION

An upgrade to the ATLAS silicon tracker cooling control system may require a change from the present $C_3F_8$ evaporant (molecular weight = 188) coolant [1] to a blend with 10-30% of the more volatile $C_2F_6$ (mw = 138). Central to this upgrade a new acoustic instrument for real-time measurement of $C_3F_8/C_2F_6$ mixture ratio and flow has been developed, exploiting the phenomenon that the sound velocity in a binary gas mixture at known temperature and pressure depends solely on the molar concentrations of its components. This instrument builds upon the technology of ultrasonic gas analysis used in Cherenkov radiation detectors since the 1980s [2] and measures the molar concentrations of the two fluorocarbon components in the recirculating exhaust vapour following the evaporative cooling of the silicon tracker.

In the custom electronics sound bursts are sent via ultrasonic transceivers◊ parallel and anti-parallel to the gas flow. A fast transit clock is started synchronously with burst transmission and stopped by over-threshold received sound pulses. Rolling average transit times in both directions, together with temperature and pressure, enter a FIFO memory and are passed to a supervisory computer via RS232 or CANbus.

Gas mixture is continuously analyzed using SCADA software implemented in PVSS-II [3], by comparing the average sound velocity in both directions with stored *velocity vs. concentration* look-up tables. These tables may be created from prior measurements in calibration mixtures or from theoretical thermodynamic calculations. Flow rates are calculated from the difference in transit time in the two directions. In future versions these calculations may be made in an on-board microcontroller.

Within the ATLAS experiment the instrument has been used for flowmetry and mixture analysis of $C_3F_8/C_2F_6$ blends and also as a sensitive detector of leaks of the present $C_3F_8$ evaporative coolant into the ATLAS pixel detector nitrogen envelope.

# MECHANICS

The mechanical envelope and ultrasonic transducer mounting are illustrated in Fig. 1. The transducers are mounted around 660 mm apart in a flanged stainless steel tube of overall length 835mm. The temperature in the tube is monitored by six NTC thermistors – (100kΩ at 25ºC) - giving an average temperature measurement uncertainty of better than ±0.3C. Pressure is monitored with a transducer having a precision better than ±15mbar.

# ELECTRONICS

The custom electronics is based on a Microchip ® dsPIC 16 bit microcontroller. This generates the 50kHz sound burst signals emitted by the transducers and includes a 40 MHz transit clock that is stopped when an amplified sound signal from a receiving transducer crosses a user-definable comparator threshold.

The HV bias for the vibrating foils of the capacitive transducers, settable in the HV range 180-360V, is generated by a DC-DC converter. When transmitting, a transducer is excited with a train of (1-8) HV square wave pulses, built using the 50kHz LV pulses from the microcontroller and the DC-DC converter output. When receiving, a transducer is biased with a flat HVDC bias and its signal passed to an AD620N amplifier followed by a comparator.

---

◊ Model 600 50kHz instrument grade ultrasonic transducer: SensComp, Inc. 36704 Commerce Rd. Livonia, MI 48150, USA

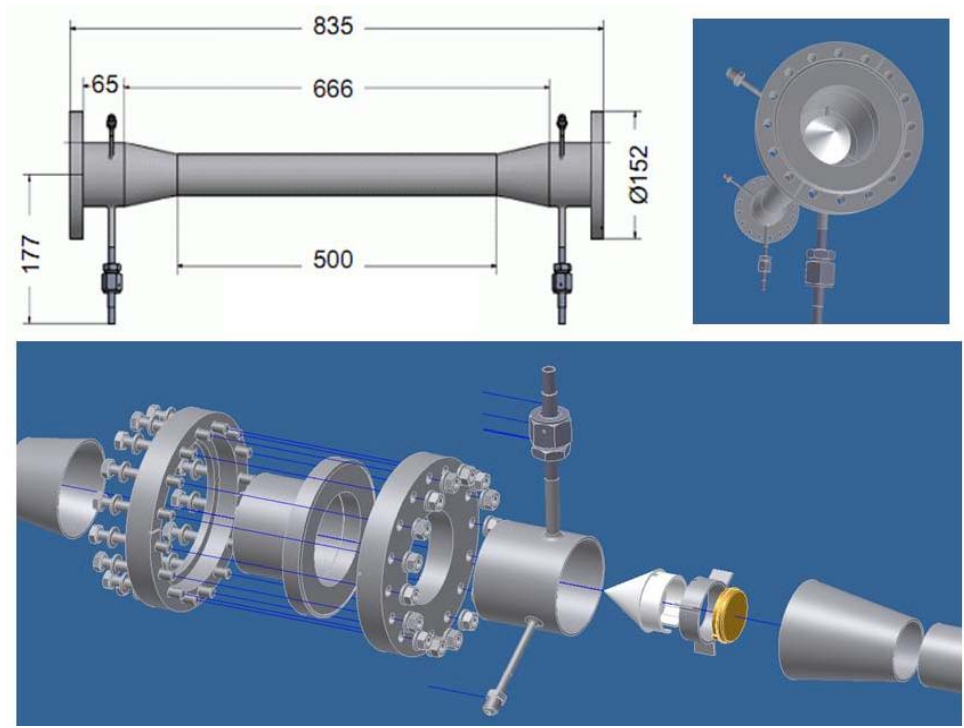

Figure 1: Views of the instrument mechanical envelope, showing an ultrasonic transducer, it's mounting and axial flow deflecting cone, together with tubes for pressure sensing and the evacuation and the injection of calibration gas.

Transit times, computed alternately in the two transmission directions are continuously entered into an internal FIFO memory. When a measuring cycle is requested by the supervisory computer a time-stamped running average from the 300 most recent transit times in each direction in the FIFO memory is output, together with the average temperature and pressure, at a rate of up to 20 averaged samples per second.

In addition to the I/O connectivity for communications, the ultrasonic transducers, pressure and temperature sensors, two (4-20 mA) analog outputs provide feedback for adjustment of the $C_3F_8/C_2F_6$ mixing ratio by the external gas mixture control system.

## CALIBRATION AND MEASUREMENTS IN PURE $C_2F_6$ AND $C_3F_8$

For high precision mixture and flow analysis the uncertainty in the sound flight distance should be minimized. It is necessary to perform a one-time transducer foil inter-distance calibration. The most convenient method is to calculate this distance from an average of measured sound transit times with the tube filled with a pure gas (or gases) having well-known sound velocity dependence on temperature and pressure. We initially made calibrations using xenon, whose sound velocity and mw (175.5 ms$^{-1}$ at 20ºC, 137 units) are closest [4], [5] to those of the fluorocarbon mixtures in the ATLAS application [1], and whose thermo-physical behaviour is that of an ideal gas. Later calibrations demonstrated sufficient precision with nitrogen and argon, which are considerably cheaper and more widely available. The average uncertainty in transducer inter-distance measured in this way is ± 0.1mm.

Multiple measurements were made in pure $C_3F_8$ and $C_2F_6$ at around 19.5°C, with pressure in the range 0.4 - 2.7bar$_{abs}$[6]; conditions expected in an acoustic vapour analyzer installed in the (superheated) vapour return path some tens of metres from the evaporative zone within the ATLAS silicon tracker. The average difference between measured sound velocities and the predictions from a PC-SAFT□ equation of state (EOS) [6], [7], [8]) was less than 0.04% in both fluids.

## SCADA & ANALYSIS SOFTWARE

The specialized software for the gas analyzer operation is coded as a standalone component in the PVSS II, v3.8 SCADA environment [3], [6]; a standard at CERN. Its main tasks include:
- vapour flow rate determination;
- sound velocity and molar vapour mixture concentration determination;
- RS232 or CANOpen communication, to start and stop the measuring cycle, and to request time-stamped bidirectional sound transit times, temperature and pressure data from the instrument FIFO memory;
- calculation and transmission of the set-points for the analog (4-20mA) output signals for $C_3F_8/C_2F_6$ ratio adjustment in the external cooling plant;
- visualization via a Graphical User Interface (GUI);
- archiving of sound transit times, velocities, flow, mixture composition, temperature and pressure into a local and/or remote data base.

The vapour flow rate is calculated from the sound transit times measured parallel, $t_{down}$, and anti-parallel, $t_{up}$, to the flow direction, according to the following algorithm:

$$t_{down} = L / (c + v), \quad t_{up} = L / (c - v) \quad (1)$$

where $v$ is the linear flow velocity (ms$^{-1}$), $c$ the speed of sound in the gas and $L$ the distance between transducers.

The gas volume flow $V$ (m$^3$s$^{-1}$) can therefore be inferred from the two transit times by:

$$V = L/2 * A * ((t_{up} - t_{down})/ t_{up} * t_{down}) \quad (2)$$

where $A$ is the internal cross sectional area of the axial flow tube between the two ultrasonic transducers (m$^2$).

The sound velocity $c$ can also be inferred from the two transit times via:

$$c = L/2 * ((t_{up} + t_{down})/ t_{up} * t_{down}) \quad (3)$$

It can be seen from eqs. (1) – (3) that knowledge of the temperature of the gas is not necessary for flowmetry.

Figure 2 shows the linearity of the ultrasonic flowmeter element of the instrument in $C_3F_8$ vapour at 20°C through comparison with a Schlumberger Delta G16 gas meter, at flows up to 230 lmin$^{-1}$ (~30 gms$^{-1}$); the maximum mass flow in the presently-available $C_3F_8/C_2F_6$ blend circulation system. The average precision is 2% of full scale.

The calculation of gas mixture molar ratio requires the use of ("*c, t, p*") look-up tables of gas mixture composition in the binary mixture to be analyzed, corresponding to sound velocities, *c*, at known temperatures, *t*, and pressures, *p*).

Look-up table data may be gathered from prior measurements in calibration mixtures or from theoretical data. Fig. 3 compares measured sound velocities in calibrated mixtures of $C_3F_8$ and $C_2F_6$ with sound velocity predictions at 19.2°C from a PC-SAFT□ EOS [7], [8] and the refrigerant-oriented extended Benedict-Webb-Rubin (BWR) EOS used in the NIST REFPROP thermo-dynamic software package [9]. The (0➔35%) $C_2F_6$ concentration range spans the region of thermodynamic interest to the ATLAS silicon tracker cooling application.

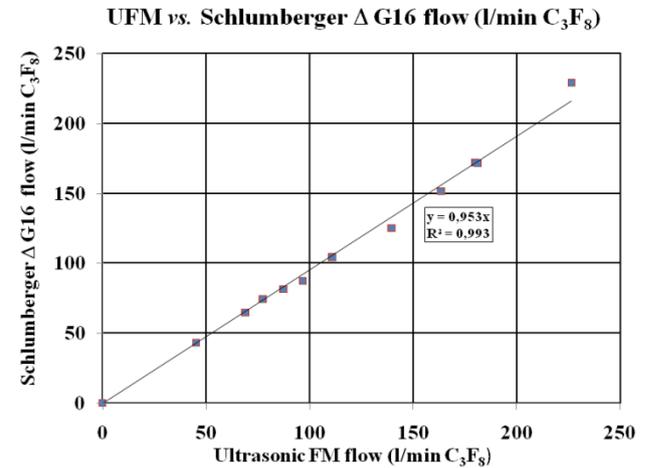

Figure 2. Ultrasonic flowmeter linearity comparison with a Schlumberger Delta G16 gas meter: $C_3F_8$ vapour.

---
□Purturbed-Chain Statistical Associating Fluid Theory

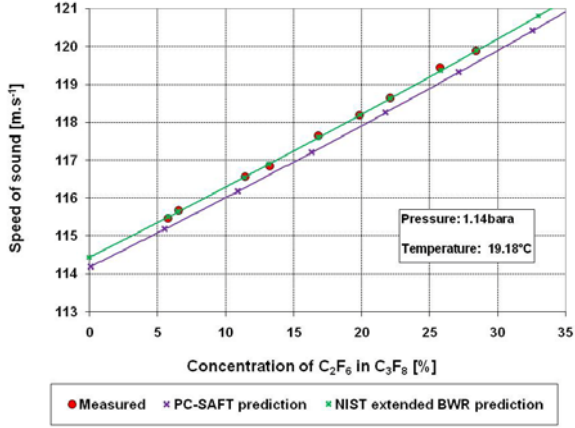

Figure 3. Comparison between measured sound velocity data and theoretical predictions in $C_3F_8$ / $C_2F_6$ mixtures.

The mixtures of $C_3F_8$ and $C_2F_6$ shown in Fig. 3 were set up by partial pressure ratio in the previously-evacuated tube, creating a molar ratio binary gas mixture. The transducer foil inter-distance had been previously established using the above-described gas calibration procedure, to a precision of ± 0.1mm.

The average difference between measured and the PC-SAFT and NIST-REFPROP predicted sound velocities in mixtures with (0➔35%) $C_2F_6$ in $C_3F_8$ were respectively 0.5% and 0.05% at pressures around 1 $bar_{abs}$ and temperatures in the range 15-25°C. It is recognised that the present version of the NIST-REFPROP [5] database is the more precise in predicting the thermophysical properties of mixtures of saturated fluorocarbons (having molecular structures of the form $C_nF_{(2n+2)}$).

The precision of mixture determination, $\delta(mix)$, at any concentration of the two components is given by;

$$\delta(mix) = \delta c/m \quad (4)$$

where $m$ is the local slope of the sound velocity/ concentration curve and $\delta c$ is the uncertainty in the sound velocity measurement - dependent on transit time resol-ution, transducer spacing and uncertainties in the measured temperature and pressure (±0.2°C, ±5mbar in this instrument) or variations between these parameters and the ($t, p$) values of the nearest ($c, t, p$) curve in the calibration database. For example, at a sound velocity of ~118 ms$^{-1}$- corresponding to a blend of 20% $C_2F_6$ in $C_3F_8$ (Fig. 3) - the combined measurement uncertainties result in a sound velocity uncertainty of 0.06 ms$^{-1}$, yielding a concentration uncertainty ~0.3% at 20%$C_2F_6$, where the slope of the velocity/concentration curve is ~0.18ms$^{-1}$%$^{-1}$.

The present software [6] uses a pre-loaded look-up table of NIST-REFPROP BWR-generated sound velocity with 0.25% granularity in $C_3F_8$/$C_2F_6$ molar mixture and covering the expected range temperature and pressures (16.2➔26.1°C, 800➔1600mbar$_{abs}$ with 0.3°C & 50mbar granularity). The algorithm calculates mixture composition by minimizing a quadratic norm, $n_i$, for each ($c_i, T_i, P_i$) table entry:

$$n_i = k_1(p_{i,\,table} - p_{running\,average})^2 + k_2(t_{i,\,table} - t_{running\,average})^2 + k_3(c_{i,\,table} - c_{running\,average})$$

(5)

where $k_{1,2,3}$ are sensitivity parameters [6] and $p$, $t$ & $c_{running\,average}$ are real-time outputs of the instrument FIFO memory.

A new software version [6] will implement a database covering a much larger $c, T, P$ range and will allow "zooming" to smaller sub-tables (O~10,000 $c, T, P$ data points), corresponding to a narrower process range - as in the present application.

In a second application related to the ATLAS evaporative cooling system, we use ultrasonic binary gas analysis to detect low level $C_3F_8$ vapour leaks into the $N_2$ environmental gas surrounding the ATLAS silicon tracker. Figure 4 compares measured sound velocities in mixtures containing up to 10% $C_3F_8$ in $N_2$ with sound velocity predictions from the PC-SAFT EOS°. A reduction in sound velocity of 0.6 m/s from the base velocity of ~351 ms$^{-1}$ was seen during a long term (> 1 year study). From the ~-12.27ms$^{-1}$%$^{-1}$ average gradient of the sound velocity-concentration curve at trace $C_3F_8$ concentrations in the range 0➔0.5% (Fig. 4) this sound velocity indicated, using

eq.(4), a $C_3F_8$ leak ingress of 0.049%, later traced to one of 204 evaporative cooling circuits into the ATLAS silicon tracker nitrogen envelope.

## CONCLUSION

We have developed a combined, real-time ultrasonic flowmeter and binary gas analyzer, whose accuracy was determined following calibration in pure reference gases by a set of measurements in $C_3F_8/C_2F_6$ blends. Sound velocity measurements were within 0.05% of the predictions of the NIST-REFPROP package, allowing mixture resolution of 0.3% in the (0-35% $C_2F_6$) concentration range of interest for the ATLAS silicon tracker.

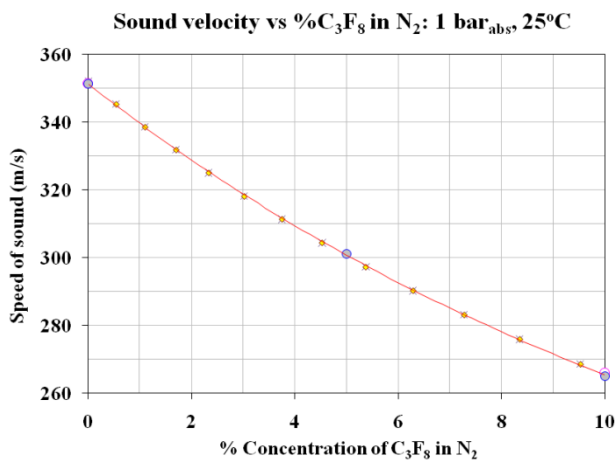

Figure 4. Comparison of sound velocity measurements (°) and PC-SAFT predictions (◊) in $C_3F_8/N_2$ mixtures.

The instrument presently analyzes vapour mixtures of $C_2F_6/C_3F_8$ and $N_2/C_3F_8$, respectively having a molecular weight difference of 50 and 160 units, the mixture resolution being seen to increase with the mw difference of the components. The instrument has applications in the analysis of hydrocarbon-air mixtures, refrigerant-air mixtures (leak detection), vapour mixtures for MOCVD semiconductor manufacture and anaesthetic gas mixtures.

## REFERENCES

[1] D. Attree et al (96 authors), "The evaporative cooling system for the ATLAS inner detector." JINST 3:P07003 (2008) 1

[2] G. Hallewell, G. Crawford, D. McShurley, G. Oxoby and R. Reif, "A sonar-based instrument for the ratiometric determination of binary gas mixtures", Nucl. Instr. & Meth A 264 (1988) 219

[3] PVSS II – Process visualization and control system, Version 3.8 (2009) ETM professional control GmbH, A-7000, Eisenstadt, Austria http://www.etm.at

[4] V. Vacek, G. Hallewell and S. Lindsay, "Velocity of sound measurements in gaseous per-fluorocarbons and their mixtures", *Fluid Phase Equilibria*, 185(2001) 305

[5] V. Vacek, G. Hallewell, S. Ilie and S. Lindsay, "Perfluorocarbons and their use in cooling systems for semiconductor particle detectors", *Fluid Phase Equilibria*, 174(2000) 191

[6] R. Bates et al (21authors), "An on-line acoustic fluorocarbon coolant mixture analyzer for the ATLAS silicon tracker"
Proc. 2nd International conference on Advancements in Nuclear Instrumentation, Measurement Methods and their Applications (ANIMMA) Ghent, Belgium 6-9 June, 2011

[7] G. Hallewell, V. Vacek and V. Vins, "Properties of saturated fluorocarbons: Experimental data and modeling using perturbed-chain-SAFT" , *Fluid Phase Equilibria* 292(1-2): 64-70 (2010)

[8] G. Hallewell, V. Vacek and M. Doubek, "Novel and simple sonar gas analyzers", *Proc 9th Asian Thermophysical Properties Conference*, Beijing, October 19–22, CD ROM Paper No.:109296: 1-6 (2010)

[9] E. Lemmon, M. Huber and M. McLinden, 'REFPROP' Standard reference database 23, version 9.0 U.S. National Institute of Standards and Technology (2010)